\documentclass[pdftex, pra, twocolumn, showpacs, floatfix, superscriptaddress, nofootinbib]{revtex4}

\usepackage{times}
\usepackage{amsfonts}
\usepackage{amssymb}
\usepackage{amsmath}
\usepackage{graphicx}
\usepackage[colorlinks=true, breaklinks=true, linkcolor=blue, citecolor=blue, urlcolor=blue]{hyperref}

\newcommand{\doi}[2]{\href{http://dx.doi.org/#1}{#2}}
\newcommand{\arxiv}[1]{\href{http://arxiv.org/abs/#1}{arXiv:#1}}
\newcommand{\urn}[1]{[\href{http://nbn-resolving.de/#1}{#1}]}

\begin{document}

\title{Quantum magnetism without lattices in strongly interacting one-dimensional spinor gases}

\author{F. Deuretzbacher}
\email{frank.deuretzbacher@itp.uni-hannover.de}
\affiliation{Institut f\"ur Theoretische Physik, Leibniz Universit\"at Hannover, Appelstrasse 2, DE-30167 Hannover, Germany}

\author{D. Becker}
\affiliation{Department of Physics, University of Basel, Klingelbergstrasse 82, CH-4056 Basel, Switzerland}

\author{J. Bjerlin}
\affiliation{Mathematical Physics, Lund Institute of Technology, SE-22100 Lund, Sweden}

\author{S. M. Reimann}
\affiliation{Mathematical Physics, Lund Institute of Technology, SE-22100 Lund, Sweden}

\author{L. Santos}
\affiliation{Institut f\"ur Theoretische Physik, Leibniz Universit\"at Hannover, Appelstrasse 2, DE-30167 Hannover, Germany}

\begin{abstract}
We show that strongly interacting multicomponent gases in one dimension realize an effective spin chain, offering an alternative simple scenario for the study of one-dimensional (1D) quantum magnetism in cold gases in the absence of an optical lattice. The spin-chain model allows for an intuitive understanding of recent experiments and for a simple calculation of relevant observables. We analyze the adiabatic preparation of antiferromagnetic and ferromagnetic ground states, and show that many-body spin states may be efficiently probed in tunneling experiments. The spin-chain model is valid for more than two components, opening the possibility of realizing SU($N$) quantum magnetism in strongly interacting 1D alkaline-earth-metal or ytterbium Fermi gases.
\end{abstract}

% 03.75.Mn: multicomponent and spinor condensates
% 75.10.Pq: spin chain models
% 67.85.Lm: degenerate Fermi gases
% 73.21.Hb: electron states and collective excitations in quantum wires

\maketitle

\section{Introduction}

Ultracold gases in optical lattices offer fascinating perspectives for the simulation of quantum magnetism, a topic of fundamental importance in condensed matter physics~\cite{Auerbach94}. Starting with the observation of superexchange in double-well systems~\cite{Trotzky08}, recent experiments are quickly advancing in the simulation of quantum and classical magnetism in optical lattices, including the creation of plaquette resonating-valence-bond states~\cite{Nascimbene12}, the simulation of a quantum Ising model using tilted lattices~\cite{Simon11, Meinert13}, the realization of classical antiferromagnetism in triangular lattices~\cite{Struck11}, and the observation of dipole-induced spin exchange in polar lattice gases~\cite{Yan13, dePaz13}. However, although short-range antiferromagnetism has been reported in dimerized lattices~\cite{Greif13}, N\'eel long-range order in two-component Fermi gases has not yet been observed, due to the very low entropy necessary in typical lattice experiments.

Strongly correlated one-dimensional (1D) systems have also attracted major attention in recent years~\cite{Cazalilla11}. Experimental developments in 1D systems are highlighted by the realization of the Tonks-Girardeau gas~\cite{Paredes04, Kinoshita04}, followed by the studies on local two- and three-body correlations~\cite{Kinoshita05, LaburtheTolra04, Haller11}, slow thermalization~\cite{Kinoshita06}, and the realization of the super-Tonks gas~\cite{Haller09}. Theoretical investigations led to several generalizations of Girardeau's Bose-Fermi mapping for spinless bosons~\cite{Girardeau60} to multicomponent systems~\cite{Girardeau07, Deuretzbacher08, Guan09, Fang11}.

Recent experiments allow for the investigation of small two-component fermionic 1D systems with a high control of particle number, spin imbalance, and interaction strength~\cite{Serwane11, Zuern12a}. These experiments have attracted considerable attention, in particular concerning the physics in the vicinity of a scattering resonance~\cite{Rontani12, Lindgren13, Bugnion13, Sowinski13, Cui13b, Volosniev13, Harshman14, Gharashi13}. For resonant interactions, the energy eigenstates show a large spin degeneracy~\cite{Deuretzbacher08, Guan09} that is lifted for finite interactions, making these systems very sensitive to temperature effects~\cite{Sowinski13} and spin segregation in the presence of magnetic-field~($B$-field) gradients~\cite{Cui13b, Cui13a}. The analytical form of the many-body wavefunction has also been addressed~\cite{Cui13b, Volosniev13, Harshman14}, although the proposed methods become very involved for large particle numbers and/or components.

\begin{figure}
\begin{center}
\includegraphics[width = \columnwidth]{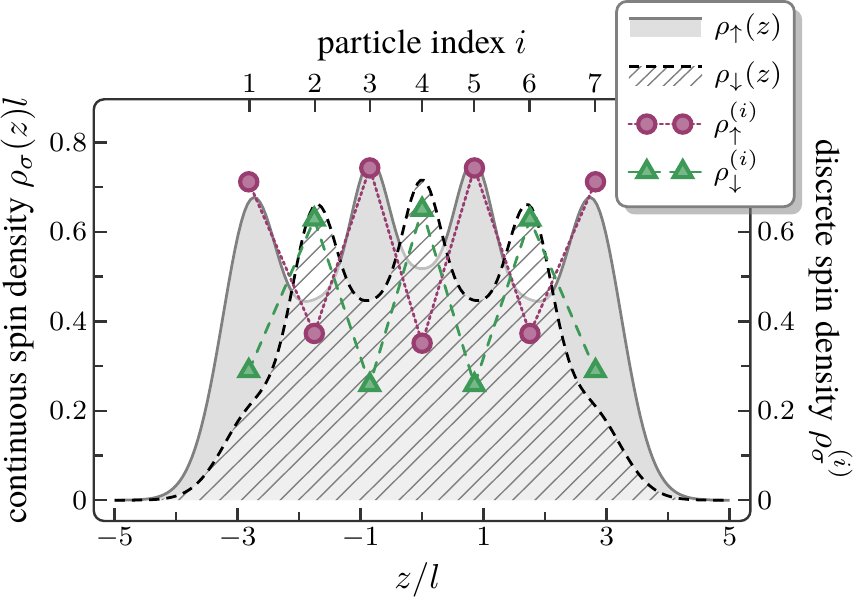}
\caption{(Color online) Continuous (experimentally measurable) spin densities ${\rho_{\uparrow, \downarrow} (z)}$ of the full model together with the discrete spin densities ${\rho_{\uparrow, \downarrow}^{(i)}}$ of the spin-chain model for seven harmonically trapped spin-1/2 fermions~${(N_\uparrow = 4, N_\downarrow = 3)}$ in the antiferromagnetic state.}
\label{fig-AF-imbalance}
\end{center}
\end{figure}

We show in this article that strongly interacting multicomponent 1D gases in the vicinity of a scattering resonance realize an effective spin chain without the need for an optical lattice. We obtain the effective spin model by combining the exact analytical solution for infinite repulsion~\cite{Deuretzbacher08} with a spin permutation model originally developed in the analysis of quantum wires~\cite{Matveev04a, Matveev04b, Fiete07}.~\footnote{The crossover to the spin-incoherent (Wigner-crystal-like) regime has been studied in Refs.~\cite{Soeffing09, Soeffing11} in the context of ultracold fermionic two-component atomic gases by analyzing the density oscillations on top of the Thomas-Fermi profile.} The resulting model significantly simplifies the calculations of the eigenfunctions and eigenenergies and may be employed for both strongly-interacting bosons and fermions. Moreover, it is applicable not only to two-component gases, but in general to multicomponent SU($N$) systems, which may be realized in alkaline-earth-metal gases and ytterbium~\cite{Gorshkov10, Taie12, Pagano14}. The specific case of spin-1/2 systems realizes an effective Heisenberg spin model, which may acquire a ferromagnetic (F) or antiferromagnetic (AF) character depending on the sign of the interparticle interactions. We analyze the dynamic creation of both an AF and a F state by making use of an exact diagonalization of the effective spin-chain model. We show finally that the properties of the spin chain may be directly measured in ongoing experiments.

\section{Noninteracting spin chain}

Multicomponent trapped Fermi or Bose systems with an infinite contact repulsion may be exactly solved~\cite{Deuretzbacher08} through a generalization of Girardeau's Bose-Fermi mapping for spinless bosons~\cite{Girardeau60}. At infinite repulsion a multicomponent 1D system behaves as a spinless Fermi gas characterized by states with a given spatial ordering of the particles. One may construct an orthonormal basis of nonsymmetric position-space sector wave functions~\cite{Deuretzbacher08}
\begin{equation}
\langle z_1, \dotsc , z_N | P \rangle = \sqrt{N!} \theta \left( z_{P(1)} , \dotsc , z_{P(N)} \right) \! A \, \psi_F ,
\end{equation}
where ${\theta (z_1, \dotsc , z_N) = 1}$ if ${z_1 \leq \dotsm \leq z_N}$ and zero otherwise, $P$ is one of the $N!$ permutations of the ordering of the $N$ particles, ${A = \prod_{i<j} \text{sgn} (z_i-z_j)}$ is the unit antisymmetric function~\cite{Girardeau60}, and $\psi_F$ is the ground state of $N$ 1D noninteracting spinless fermions. The eigenfunctions of multicomponent Bose and Fermi systems are obtained via the map~\cite{Deuretzbacher08}
\begin{equation} \label{map}
W_\pm | \chi \rangle = \sqrt{N!} S_\pm \left( | \text{id} \rangle | \chi \rangle \right) ,
\end{equation}
where ${| \chi \rangle = \sum_{m_1, \dotsc, m_N} c_{m_1, \dotsc, m_N} | m_1, \dotsc, m_N \rangle}$ is an arbitrary $N$-particle spin function, ${S_\pm = (1/N!) \sum_P (\pm 1)^P P}$ is the (anti)symmetrization operator, and ${| \text{id} \rangle}$ is the sector wave function corresponding to the identical permutation.~\footnote{The map~(\ref{map}) can be easily extended to states with excited spatial degrees of freedom by replacing the ground state $\psi_F$ in the sector wave functions ${| P \rangle}$ by the $i$th excited state $\psi_F^{(i)}$.} An important consequence of the bijective character of the map~(\ref{map}) is that the system is uniquely determined by the spin function ${| \chi \rangle}$. In particular, the density distribution of the $m$th component is given by~\cite{Deuretzbacher08}
\begin{equation}
\rho_m (z) = \sum_i \rho_m^{(i)} \rho^{(i)}(z)
\end{equation}
with the probability that the magnetization of the $i$th spin equals $m$,
\begin{equation}
\rho_m^{(i)} = \sum_{m_1, \dotsc, m_N} \bigl| \langle m_1, \dotsc, m_N | \chi \rangle \bigr|^2 \delta_{m, m_i} ,
\end{equation}
and the probability to find the $i$th particle (with whatever spin) at position~$z$,
\begin{equation}
\rho^{(i)}(z) = N! \int d z_1 \dotsi d z_N \delta (z-z_i) \theta (z_1, \dotsc , z_N) \bigl| \psi_F \bigr|^2 .
\end{equation}
The continuous spin density ${\rho_m (z)}$ is hence fully characterized by the $N$-tuple ${\bigl( \rho_m^{(1)}, \dotsc, \rho_m^{(N)} \bigr)}$, as illustrated in Fig.~\ref{fig-AF-imbalance}. The system thus reduces to a spin-chain model.

\section{Spin-spin interactions}

In the limit of infinite repulsion, ${1/g = 0}$~(with the interaction strength $g$), the spin chain is noninteracting, since all states of the ground-state multiplet are degenerate. This is no longer the case when ${1/g \neq 0}$. In the vicinity of a scattering resonance the effective theory for finite interactions may be evaluated to lowest order in ${1/g}$ by means of degenerate perturbation theory. The effective interaction Hamiltonian of the spin chain reads (see Appendix~\ref{app-Hspin} for the derivation)~\footnote{For particles on a ring, one has to replace ${N-1}$ by $N$ in Eq.~(\ref{Hspin}) and $P_{N,N+1}$ has to be replaced by $P_{N,1}$.}
\begin{equation} \label{Hspin}
H_s = \left( E_F - \sum_{i=1}^{N-1} J_i \right) \openone \pm \sum_{i=1}^{N-1} J_i P_{i,i+1} ,
\end{equation}
where $P_{i,i+1}$ denotes the permutation of the spin of neighboring particles, the $+$ ($-$) sign applies to fermions (bosons), and the nearest-neighbor exchange constants are given by
\begin{equation} \label{J}
J_i = \frac{N! \hbar^4}{m^2 g} \int dz_1 \dotsi dz_N \delta (z_i-z_{i+1}) \theta (z_1, \dotsc, z_N) \left| \frac{\partial \psi_F}{\partial z_i} \right|^2 \! .
\end{equation}
The exact calculation of the exchange constants $J_i$ requires the solution of multidimensional integrals of growing complexity with increasing $N$, which is in practice possible only for small $N$.~\footnote{See the second version of Ref.~\cite{Volosniev13}.} Fortunately, an accurate approximation of the exchange constants, which becomes even more accurate for growing $N$, is provided by the expression
\begin{equation} \label{Japprox}
J_i = \frac{\hbar^4 \pi^2 n_\text{TF}^3(z_i)}{3 m^2 g} ,
\end{equation}
where $n_\text{TF}$ is the Thomas-Fermi~(TF) profile of the density and $z_i$ is the center of mass of the $i$th and $(i+1)$th particle density, ${\rho^{(i)}(z)}$ and ${\rho^{(i+1)}(z)}$~(see Appendix~\ref{app-J}). Expression~(\ref{Japprox}) follows from the nearest-neighbor exchange of the homogeneous system with periodic boundary conditions in the thermodynamic limit~\cite{Matveev08} combined with a local density approximation~(LDA). Appendix~\ref{app-J} shows a comparison between exchange constants obtained from Eqs.~(\ref{J}) and~(\ref{Japprox}) for up to six harmonically trapped particles, confirming that, as mentioned above, the agreement becomes better for growing $N$.

The diagonalization of the spin Hamiltonian~(\ref{Hspin}) in combination with the map~(\ref{map}) allows for a simple calculation of the eigenstates of trapped strongly interacting multicomponent bosons or fermions.~\footnote{For three spin-1/2 fermions ${(N_\uparrow = 2, N_\downarrow = 1)}$ our results agree with those presented in the first version of Ref.~\cite{Volosniev13}. The position dependence of the nearest-neighbor exchange constants~(\ref{J}) was recently noted in the second version of Ref.~\cite{Volosniev13}.} This means that the spin distribution, and hence the whole atom distribution in the trap, is determined by a spin permutation Hamiltonian~(Sutherland model~\cite{Sutherland75}). In the case of spin-1/2 particles we have ${P_{i,i+1} = (\vec \sigma^{(i)} \cdot \vec \sigma^{(i+1)} + 1)/2}$ with the Pauli vector $\vec \sigma$. Two-component gases therefore realize an effective Heisenberg Hamiltonian. The Heisenberg Hamiltonian coincides with that introduced in the analysis of the conductance of quantum wires~\cite{Matveev04a, Matveev04b, Fiete07} and of spectral functions of spin-1/2 1D bosons~\cite{Matveev08}. The effective spin model is consistent with Bethe-ansatz results for spin-1/2 bosons~\cite{Guan07} and fermions~\cite{Lee12}. The validity of the spin-chain model is restricted to the (super-)Tonks regime, where ${|1/g|}$ is small (see Appendix~\ref{app-validity} for a comparison with a numerical exact diagonalization of the full Hamiltonian).

\section{Spin order}

\begin{figure}
\begin{center}
\includegraphics[width = \columnwidth]{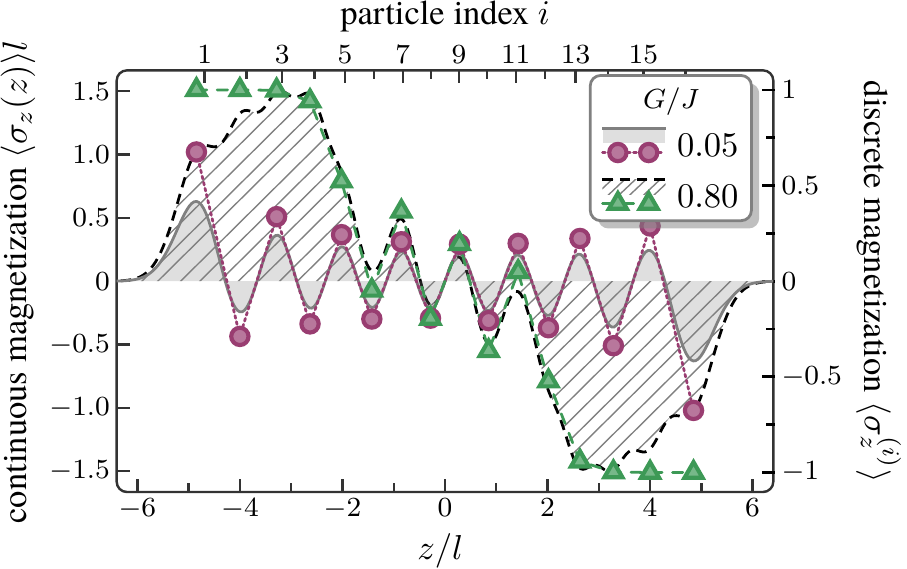}
\caption{(Color online) Magnetization of a spin-balanced AF spin chain consisting of 16 harmonically trapped particles for ${G/J = 0.05}$ and $0.8$ ($G$ is the $B$-field gradient and ${J = \sum_i J_i / (N-1)}$ is the average nearest-neighbor exchange). The symbols (shaded curves) denote the discrete (continuous) distributions.}
\label{fig-AF-gradient}
\end{center}
\end{figure}

In the following we focus on the specific case of spin-1/2 gases, which is of direct relevance for ongoing experiments~\cite{Serwane11, Zuern12a}. Equation~(\ref{J}) (${J_i \propto 1/g}$) implies that the sign of the $J_i$ can be tuned by means of a scattering resonance~\cite{Zuern12a}. The spin interaction is F for ${g<0}$~(${g>0}$) and AF for ${g>0}$~(${g<0}$) for fermions~(bosons). Although spin-spin correlations would clearly show the (anti)ferromagnetic character of the interactions, for both F and AF couplings, the local magnetization ${\bigl\langle \sigma_z^{(i)} \bigr\rangle}$ is zero for all particle positions in the ground state due to SU(2) symmetry. As a result, the density distributions of both spin components will be identical. This symmetry may be broken by a small population imbalance~(Fig.~\ref{fig-AF-imbalance}; see also Ref.~\cite{Guan09}) or by a spin-dependent external potential, such as a $B$-field gradient~(Fig.~\ref{fig-AF-gradient}). Such a gradient adds to the effective spin interaction Hamiltonian~(\ref{Hspin}) a term ${V_G = (G/l) \sum_i \langle z \rangle_i \sigma_z^{(i)}}$ with ${\langle z \rangle_i = \int dz z \rho^{(i)}(z)}$ and the oscillator length $l$~(Appendix~\ref{app-gradient}). A small ${G/J}$ [${J = \sum_i J_i / (N-1)}$ is the average nearest-neighbor exchange constant] results in an alternating distribution of the two components marking the AF order. In contrast, when ${G/J}$ is sufficiently large the system experiences spin segregation. Since ${|J|}$ is very small at the resonance such segregation may occur for rather weak $B$-field gradients~\cite{Cui13b}. We stress, however, that this spin segregation occurs even for AF interactions, and does not mark an AF-F transition, being rather a Stern-Gerlach- (SG-)like separation of the components.

\section{State preparation}

In contrast to experiments in optical lattices, where spin ground states are exceedingly difficult to prepare, the realization of ground states of effective 1D spin chains may be accomplished in a surprisingly simple way (for the AF regime) in ongoing experiments on strongly interacting spin-1/2 fermions~\cite{Serwane11, Zuern12a}. The system is first prepared in the spin-singlet ground state of the noninteracting system.~\footnote{Temperature effects may be significant if the sample is cooled down close to the resonance~\cite{Sowinski13}, and in particular if ${k_B T > NJ}$ the system becomes a spin-incoherent Luttinger liquid~\cite{Fiete07}. This is however not relevant in typical experiments, since the initial sample is produced far from resonance.} The interaction strength $g$ is then ramped up by means of a scattering resonance into the regime of large ${g>0}$~(Tonks regime). Due to spin conservation the noninteracting ground state evolves into an AF spin chain. As discussed below, the AF order may be easily revealed in ongoing tunneling experiments using imbalanced mixtures.

\begin{figure}
\begin{center}
\includegraphics[width = 0.9 \columnwidth]{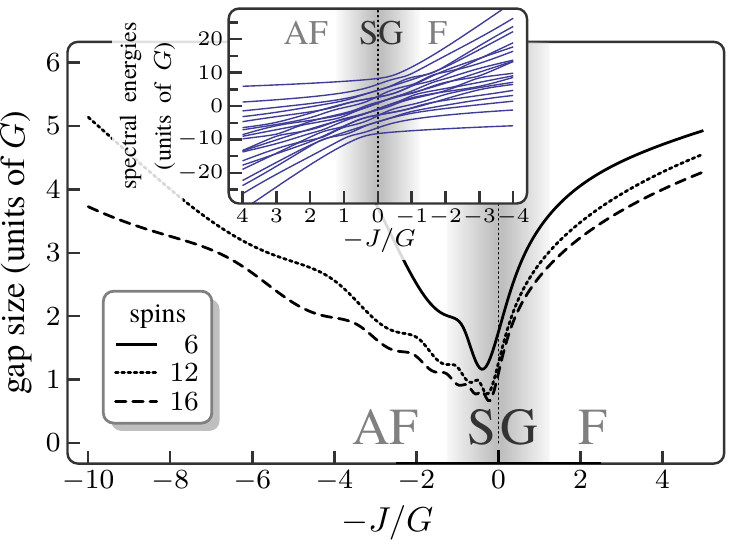}
\caption{(Color online) Gap between the ground and first excited state of harmonically trapped spin-balanced spin-1/2 fermions for nonzero gradients (${G \neq 0}$) around the resonance. While spin interactions dominate in the AF and F regimes, the $B$-field gradient dominates in the gray-shaded Stern-Gerlach (SG) regime, characterized by SG-like spin segregation. Inset: Spectrum of six spin-balanced spin-1/2 fermions as a function of ${-J/G}$.}
\label{fig-gap}
\end{center}
\end{figure}

The preparation of the spin ground state is more involved if it demands a sweep through the scattering resonance. If the system is driven across ${J=0}$, the ground state of the Tonks regime becomes the highest excited state of the super-Tonks regime~(${g<0}$),~\footnote{For ${g<0}$ the lowest energy corresponds actually to molecular states, but these states cannot be reached in a sweep.} which is preserved due to spin conservation~\cite{Cui13b}. A spin-dependent external potential, such as, e.g., a $B$-field gradient, violates spin conservation, lifting the spin degeneracy at ${J=0}$~\cite{Sowinski13, Cui13b}~(inset of Fig.~\ref{fig-gap}). In particular, the AF ground state for ${g>0}$ may be adiabatically transformed into the F ground state for ${g<0}$ due to the avoided crossing opened by the $B$-field gradient, as suggested in Ref.~\cite{Cui13b}. We employ below the spin model to analyze the conditions for the adiabatic sweep in the presence of a $B$-field gradient.~\footnote{Experiments performed by Jochim and co-workers employ a scattering resonance at 783\,G, well within the Paschen-Back regime, in which the energies of the employed states ${| F = 1/2, m_F = \pm 1/2 \rangle}$ show the same $B$-field dependence. As a result a $B$-field gradient does not lift the degeneracy at ${1/g=0}$, precluding in this experiment the use of sweeps to reach the F ground state in the super-Tonks regime.}

The gap $\Delta$ between the ground and first excited state is particularly relevant, since adiabaticity requires that ${|J/G|}$ is varied much more slowly than ${\hbar / \Delta}$. We have calculated the gap as a function of ${-J/G}$ for up to 16 spin-balanced spin-1/2 fermions by means of an exact diagonalization of the effective spin Hamiltonian.~\footnote{We note in passing that the exact diagonalization of the original Hamiltonian may be accomplished only for very few particles ${N \leq 5}$~\cite{Bugnion13, Sowinski13, Gharashi13} for (quasi)balanced mixtures, whereas the spin-chain model allows for exact diagonalizations of rather large samples ${N \leq 20}$ (and the treatment of even much larger $N$ using, e.g., density-matrix renormalization-group techniques). For the particularly favorable case of ${(N_\uparrow = N-1, N_\downarrow = 1)}$ systems, up to ${N \leq 7}$ particles have been calculated using the original Hamiltonian~\cite{Lindgren13}, whereas $N$ up to several thousands can be handled using the spin-chain model.} Figure~\ref{fig-gap} shows that the minimal gap ${\Delta_\text{min} \approx G}$ is reached in the Tonks regime (${-J/G \lesssim 0}$), and that $\Delta_\text{min}$ decreases slowly with larger $N$. Also note that the region where ${\Delta \simeq \Delta_\text{min}}$ increases with increasing $N$. This implies that an adiabatic sweep becomes more involved for larger $N$, since ${-J/G}$ has to be increased much more slowly than ${\hbar / \Delta_\text{min}}$ in an increasing region of the Tonks regime.

\begin{figure}
\begin{center}
\includegraphics[width = 0.9 \columnwidth]{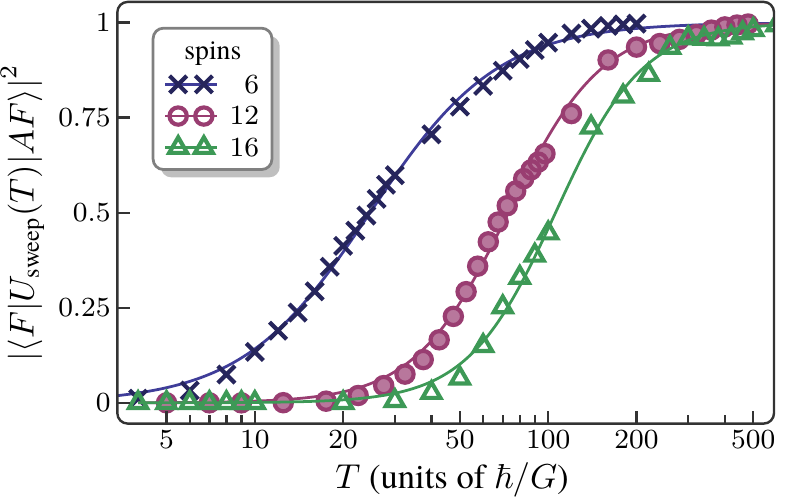}
\caption{(Color online) Overlap between the F ground state of harmonically trapped spin-balanced spin-1/2 fermions expected for ${-J/G = 10}$ and the state obtained after a linear sweep across the resonance starting with the AF ground state for ${-J/G = -10}$.}
\label{fig-sweep}
\end{center}
\end{figure}

We have perfomed exact time-dependent simulations with linear sweeps ${-J(t)/G = -10 (1-2t/T)}$ for different values of the sweeping time $T$. The initial and final values satisfy ${|J/G| \gg 1}$, and hence any final F state is maintained by F interactions and not by a SG-like spin segregation~\cite{Cui13b}. We have calculated the overlap between the state after the sweep and the F ground state. As expected, adiabaticity demands a slower sweep for larger $N$. Figure~\ref{fig-sweep} shows that in order to reach the F ground state of the super-Tonks regime with ${\simeq 100\%}$ fidelity, the sweep must fulfill ${v \equiv \partial{|J/G|} / \partial t < v_c \simeq 0.07 G / \hbar}$ in the vicinity of the resonance. This corresponds to ${T > 300 \hbar / G}$ in Fig.~\ref{fig-sweep}. We note that, although we have chosen a linear sweep for simplicity, the ramp may be much faster far from the resonance, as long as ${v < v_c}$ in the region of the minimal gap. Once the F state is reached at ${|J| \gg G}$, the $B$-field gradient may be removed. Note again that due to SU(2) symmetry the final F state does not show spin segregation if ${|J/G| \gg 1}$.

\section{State detection}

As discussed above, ${\bigl\langle \sigma_z^{(i)} \bigr\rangle}$ is mapped on the densities of the spin components. The AF or F spin ordering of the spin chain may therefore be directly probed in imbalanced mixtures by means of {\it in situ} imaging, which is however challenging in tightly confined samples. An alternative way of probing the spin order is provided by the tunneling techniques recently developed by Jochim and co-workers~\cite{Serwane11, Zuern12a}. A tight dipole trap is combined with a $B$-field gradient, which lowers the potential barrier at the right-hand side of the trap. The tunneling through this barrier may be controlled by carefully varying the $B$-field gradient. The barrier height and the waiting time may be chosen such that only one fermion can tunnel. Afterwards the spin orientation of this fermion is detected. Within the spin-chain picture only the rightmost particle can tunnel, since the particles cannot interchange their positions. The spin-chain picture hence provides a definite prediction about the spin orientation of the outcoupled fermion. We illustrate this for the specific case of a ${(N_\uparrow = 2, N_\downarrow = 1)}$ system in the Tonks~(AF) regime for ${1/g \rightarrow 0}$. The spin model provides the AF ground state ${| 0 \rangle \equiv ( | \uparrow, \uparrow, \downarrow \rangle - 2 | \uparrow, \downarrow, \uparrow \rangle + | \downarrow, \uparrow, \uparrow \rangle ) / \sqrt{6}}$. The probability of outcoupling a single down spin is therefore ${| \langle \uparrow, \uparrow, \downarrow | 0 \rangle|^2 \simeq 16.7\%}$, in very good agreement with the experiment~\cite{Zuern12b}.~\footnote{A similar result~(${\simeq 20\%}$) was predicted in the first version of Ref.~\cite{Volosniev13}. This result was recently refined~(${\simeq 16.7\%}$) in the second version of Ref.~\cite{Volosniev13}, in excellent agreement with our result obtained from the spin-chain model.} By contrast, if the system is prepared in the first excited state, ${| 1 \rangle \equiv ( | \uparrow, \uparrow, \downarrow \rangle - | \downarrow, \uparrow, \uparrow \rangle ) / \sqrt{2}}$, the probability is ${| \langle \uparrow, \uparrow, \downarrow | 1 \rangle|^2 \simeq 50\%}$ and in the F highest excited state, ${| 2 \rangle \equiv ( | \uparrow, \uparrow, \downarrow \rangle + | \uparrow, \downarrow, \uparrow \rangle + | \downarrow, \uparrow, \uparrow \rangle ) / \sqrt{3}}$, we get ${| \langle \uparrow, \uparrow, \downarrow | 2 \rangle|^2 \simeq 33.3\%}$. A similar simple calculation predicts the probabilities ${5.1\%}$ and ${1.5\%}$ for the AF ground states of ${(3,1)}$ and ${(4,1)}$ systems,~\footnote{The same results were recently presented in the second version of Ref.~\cite{Volosniev13}.} respectively, and much larger probabilities for the corresponding excited states. This measurement may hence clearly reveal the AF ground state.

\begin{figure}
\begin{center}
\includegraphics[width = \columnwidth]{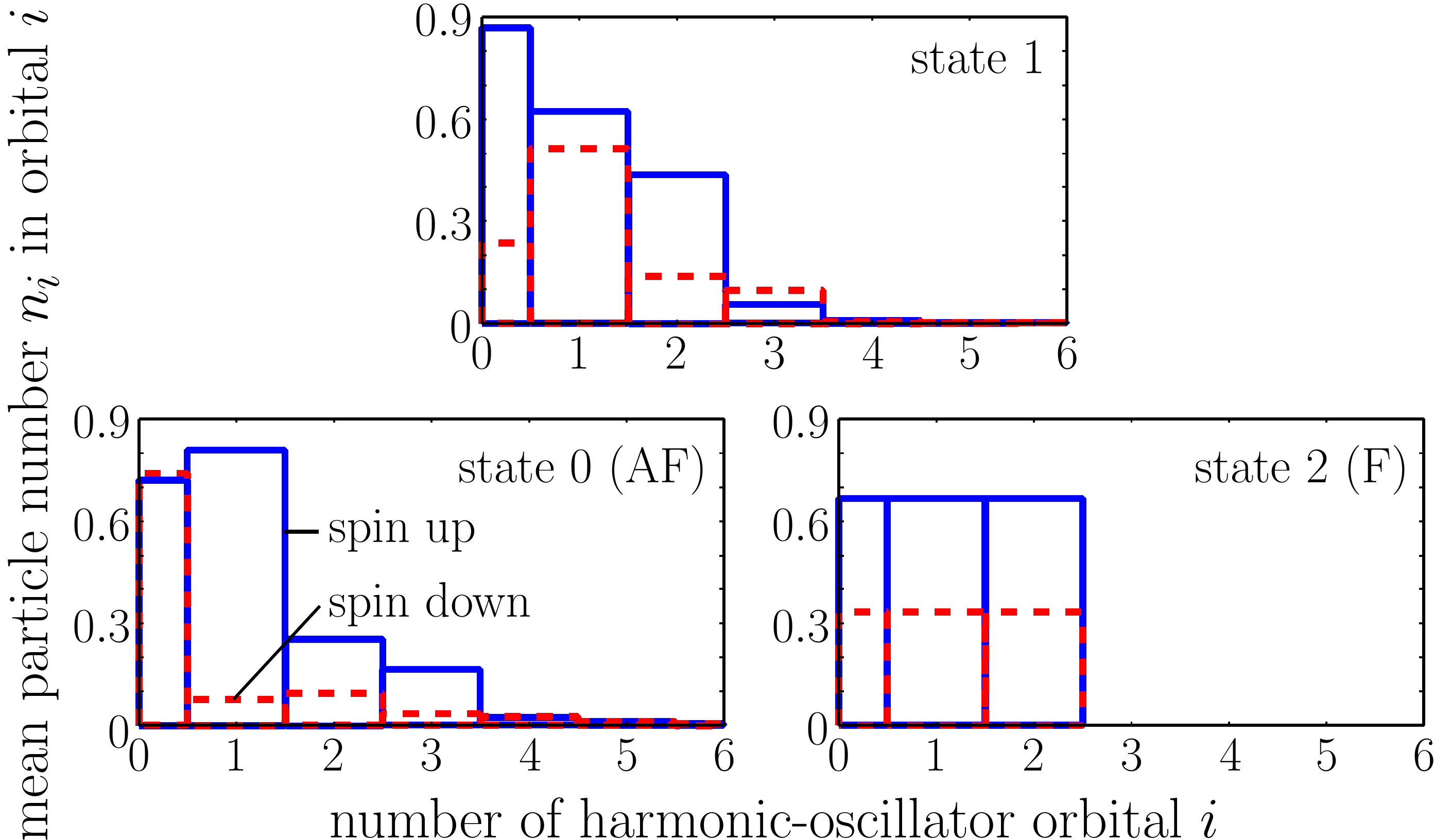}
\caption{(Color online) Mean occupation ${\langle n_i \rangle}$ of the harmonic-trap levels for the ${(N_\uparrow = 2, N_\downarrow = 1)}$ system in the Tonks regime ${[g / (\hbar \omega l) = 25]}$ for the states ${| 0 \rangle}$, ${| 1 \rangle}$, and ${| 2 \rangle}$ (see text) of the ground-state multiplet.}
\label{fig-occupancies-uud}
\end{center}
\end{figure}

Tunneling experiments may also be employed to measure the occupation-number distribution among the trap levels. First, the spin-up~(-down) fermions are removed with a resonant light pulse, and afterwards the occupancies of the remaining spin-down~(-up) fermions are probed using the tunneling technique~\cite{Zuern13}. Each spin state is linked to a particular occupation number distribution of the spin components among the trap levels~(Fig.~\ref{fig-occupancies-uud}). One may hence utilize this information as a fingerprint of the state of the spin chain [see Appendix~\ref{app-occ} for the discussion of the ${(N_\uparrow = 3, N_\downarrow = 2)}$ five-fermion system].

\section{Experimental requirements}

As mentioned above, the creation of the AF state with fermions does not require crossing the scattering resonance. Due to spin conservation it may be created by increasing ${g>0}$ starting with the noninteracting (spin-singlet) ground state~(initial particle-hole excitations will be mapped on spin excitations of the AF chain). Realizing and probing the ground state of the 1D AF spin chain requires hence the deterministic preparation of noninteracting ground states, together with a good isolation from the environment, single-atom detection, precise control of $g$, and quasi-1D confinement. These conditions are already met in ongoing experiments on degenerate lithium-6 atoms~\cite{Serwane11, Zuern12a, Zuern12b}. These experiments allow for the preparation of the noninteracting ground state with a fidelity of $98\%$ per atom, for a precise control of the spin imbalance, for the modification of $g$ using a confinement-induced resonance, and for single-atom detection with near unit fidelity. The system is very well isolated, with a lifetime of the two-particle ground state of 1 min. These conditions result in an effective spin temperature of zero, even for strong interactions, and hence this setup constitutes an optimal scenario for the realization of AF chains.~\footnote{For moderate trap aspect ratios, like those used by Jochim and co-workers~\cite{Serwane11, Zuern12a}, the coupling of center-of-mass and relative motion~\cite{Sala13} may result in the formation of molecules in the Tonks regime~\cite{Zuern12b, Gharashi12}. This problem can be avoided using larger trap aspect ratios.} Although the experiments of Refs.~\cite{Serwane11, Zuern12a, Zuern12b} are currently limited to small samples~($N<10$), much larger ones, and hence longer spin chains, may be achieved in similar experiments by increasing the trap aspect ratio~(currently $1:10$) and improving the fidelity in the preparation of the noninteracting ground state.

\section{Summary}

Strongly interacting multicomponent 1D gases in the vicinity of a scattering resonance realize a 1D spin chain, providing a novel scenario for the study of quantum magnetism alternative to atoms in 1D optical lattices and ion traps~\cite{Blatt12}. This alternative scenario, which avoids the inherent heating associated with an optical lattice, opens the possibility of creating an AF state from a noninteracting singlet state by simply increasing the interaction strength. Moreover, the effective spin-chain model provides a simple and intuitive understanding of recent experiments, allows for a very simple calculation of relevant observables, and enables numerical simulations of the statics and dynamics of much larger samples than the original model.

Although we have focused mainly on the spin-1/2 case, the spin-chain picture is equally valid for higher spins. Interestingly, strongly interacting alkaline-earth-metal or ytterbium Fermi gases realize an SU($N$) Sutherland model. In particular, a spin-3/2 system would realize an SU(4) exchange Hamiltonian, which is of relevance in spin-orbital models of transition-metal oxides. The ground state of this system is a spin liquid, since magnetic order is suppressed due to orbital effects~\cite{Yamashita98}. Moreover, magnetic-field gradients may be employed to prepare nontrivial initial spin states (e.g., helical states), and to rotate individual spins in combination with radio-frequency fields. This would allow for the study of the subsequent dynamics of the out-of-equilibrium 1D spin chain. Experiments on 1D strongly interacting multicomponent Fermi gases hence open a fascinating alternative scenario for the simulation of 1D quantum spin chains in cold gases.

\section*{\uppercase{Acknowledgments}}

We thank G.~Z\"urn, S.~Jochim, T.~Lompe, S.~Murmann, J.~C.~Cremon, N.~L.~Harshman, S.~Eggert, C.~Klempt, M.~Valiente, and L.~H.~Kristinsd\'ottir for helpful discussions. This work was supported by the Cluster of Excellence QUEST, the German-Israeli foundation, the Swiss SNF, the NCCR Quantum Science and Technology, the Swedish Research Council, and the Nanometer Structure Consortium at Lund University.

\begin{appendix}

\section{Effective interaction Hamiltonian}
\label{app-Hspin}

We derive in this appendix the effective Hamiltonian for interactions between nearest-neighboring spins of the spin chain in the vicinity of the point ${1/g=0}$. It has been shown in Ref.~\cite{Deuretzbacher08} that the spin chain is noninteracting at ${1/g=0}$ and highly degenerate due to the large number of possible spin configurations. This degeneracy is lifted away from ${1/g=0}$ but the eigenstates at ${1/g \approx 0}$ are still very well approximated by particular superpositions of the eigenstates at ${1/g=0}$, as shown in Fig.~1 of Ref.~\cite{Deuretzbacher08}. This suggests determination of the superpositions by performing a degenerate perturbative calculation to lowest order in ${1/g}$. In the following we derive the effective spin Hamiltonian, which leads to the desired superposition of spin states in the vicinity of ${1/g=0}$.

We construct for small ${1/g}$ the $g$-dependent sector wave functions
\begin{eqnarray}
\mspace{-30mu} \langle z_1, \dotsc , z_N | P^{(g)} \rangle & = & \psi_P^{(g)} (z_1, \dotsc , z_N) \nonumber \\
& = & \sqrt{N!} \theta \left( z_{P(1)} , \dotsc , z_{P(N)} \right) \psi_B^{(g)}
\end{eqnarray}
with ${\theta (z_1, \dotsc , z_N) = 1}$ if ${z_1 \leq \dotsb \leq z_N}$ and zero otherwise, $P$ is one of the $N!$ permutations of the ordering of the $N$ particles, and $\psi_B^{(g)}$ is the ground state of $N$ 1D spinless $\delta$-interacting bosons. They converge in the limit ${1/g \rightarrow 0}$ towards the usual $g$-independent sector wave functions
\begin{eqnarray}
\mspace{-30mu} \langle z_1, \dotsc , z_N | P \rangle & = & \psi_P (z_1, \dotsc , z_N) \nonumber \\
& = & \sqrt{N!} \theta \left( z_{P(1)} , \dotsc , z_{P(N)} \right) \! A \, \psi_F
\end{eqnarray}
with the unit antisymmetric function ${A = \prod_{i<j} \text{sgn} (z_i-z_j)}$ and the ground state of $N$ 1D spinless noninteracting fermions $\psi_F$. We approximate the exact wave functions in the vicinity of ${1/g=0}$ by
\begin{equation} \label{map-g}
W_\pm^{(g)} | \chi \rangle = \sqrt{N!} S_\pm \left( | \text{id}^{(g)} \rangle | \chi \rangle \right) ,
\end{equation}
where ${| \chi \rangle = \sum_{m_1, \dotsc, m_N} c_{m_1, \dotsc, m_N} | m_1, \dotsc, m_N \rangle}$ is an arbitrary $N$-particle spin function, ${S_\pm = (1/N!) \sum_P (\pm 1)^P P}$ is the (anti)symmetrization operator, and ${| \text{id}^{(g)} \rangle}$ is the sector wave function corresponding to the identical permutation. Our goal is to calculate the matrix elements
\begin{equation} \label{Hij-1}
\langle m_1, \dotsc | \left( W_\pm^{(g)} \right)^\dagger H \, W_\pm^{(g)} | m_1', \dotsc \rangle
\end{equation}
of the full many-body Hamiltonian in the vicinity of ${1/g=0}$. Inserting Eq.~(\ref{map-g}) into Eq.~(\ref{Hij-1}) and using ${S_\pm^\dagger = S_\pm}$, ${[ H, S_\pm ] = 0}$, and ${S_\pm^2 = S_\pm}$, we get
\begin{eqnarray} \label{Hij-2}
& & \mspace{-50mu} \langle m_1, \dotsc | \left( W_\pm^{(g)} \right)^\dagger H \, W_\pm^{(g)} | m_1', \dotsc \rangle \nonumber \\
& & \mspace{-50mu} = \sum_P (\pm 1)^P \langle m_1, \dotsc | \langle \text{id}^{(g)} | H | P^{(g)} \rangle | m_{P^{-1}(1)}', \dotsc \rangle .
\end{eqnarray}
Next we evaluate the matrix elements ${\langle \text{id}^{(g)} | H | P^{(g)} \rangle}$. The first two terms of the Taylor series of these matrix elements around ${1/g=0}$ are given by
\begin{eqnarray} \label{Hij-3}
& & \langle \text{id}^{(g)} | H | P^{(g)} \rangle = \lim_{1/g \rightarrow 0} \left( \langle \text{id}^{(g)} | H | P^{(g)} \rangle \right) \nonumber \\
& & + \frac{1}{g} \lim_{1/g \rightarrow 0} \left( \langle \text{id}^{(g)} | \frac{d H}{d (1/g)} | P^{(g)} \rangle \right) \nonumber \\
& & = E_F \delta_{P, \text{id}} - \frac{1}{g} \lim_{g \rightarrow +\infty} \left( g^2 \langle \text{id}^{(g)} | \frac{d H}{d g} | P^{(g)} \rangle \right) .
\end{eqnarray}
Here we used ${H | P^{(g)} \rangle = E^{(g)} | P^{(g)} \rangle}$ and $d \langle \text{id}^{(g)} | P^{(g)} \rangle / d g = 0$. The Hamiltonian of the multicomponent particles reads
\begin{equation} \label{H}
H = \sum_i \left[ -\frac{\hbar^2}{2m} \frac{\partial^2}{\partial z_i^2} + V(z_i) \right] + g \sum_{i<j} \delta(z_i-z_j)
\end{equation}
and therefore
\begin{eqnarray} \label{dHijdg-1}
& & \mspace{-20mu} \lim_{g \rightarrow +\infty} \left( g^2 \langle \text{id}^{(g)} | \frac{d H}{d g} | P^{(g)} \rangle \right) \nonumber \\
& & \mspace{-20mu} = \sum_{i<j} \lim_{g \rightarrow +\infty} \left[ g^2 \int dz_1 \dotsi dz_N \, \delta(z_i-z_j) \left( \psi_\text{id}^{(g)} \right)^{\!\!*} \psi_P^{(g)} \right] . \nonumber \\
& &
\end{eqnarray}
Most integrals are zero, since the corresponding domain of integration has zero volume; hence
\begin{eqnarray}
& & \mspace{-50mu} \int dz_1 \dotsi dz_N \delta (z_i-z_j) \nonumber \\
& & \mspace{-50mu} \mspace{19mu} \times \theta (z_1, \dotsc, z_N) \theta (z_{P(1)}, \dotsc, z_{P(N)}) \dotsm \nonumber \\
& & \mspace{-50mu} = \delta_{j,i+1} \Bigl( \delta_{P, \text{id}} + \delta_{P, P_{i,i+1}} \Bigr) \nonumber \\
& & \mspace{-50mu} \mspace{17mu} \times \int dz_1 \dotsi dz_N \delta (z_i-z_{i+1}) \theta (z_1, \dotsc, z_N) \mspace{2mu} \dotsm \, .
\end{eqnarray}
Moreover, the limit in Eq.~(\ref{dHijdg-1}) can be performed, since the average local correlation function of spinless bosons,
\begin{equation}
\int dz_1 \dotsi dz_N \sum_{i<j} \delta(z_i-z_j) \bigl| \psi_B^{(g)} \bigr|^2 ,
\end{equation}
decreases proportionally to ${1/g^2}$ in the limit of large $g$.~\footnote{This has been shown for homogeneous systems in the thermodynamic limit~\cite{Gangardt03} and it also follows from the solution of two harmonically trapped particles~\cite{Busch98}. It is hence natural to assume that this property holds true for an arbitrary number of particles in any confinement.} Using the boundary condition
\begin{eqnarray}
& & \left( \frac{\partial}{\partial z_i} - \frac{\partial}{\partial z_j} \right) \psi \big|_{z_i=z_j+} - \left( \frac{\partial}{\partial z_i} - \frac{\partial}{\partial z_j} \right) \psi \big|_{z_i=z_j-} \nonumber \\
& & = \frac{2 m g}{\hbar^2} \psi \big|_{z_i=z_j} ,
\end{eqnarray}
which is imposed by the $\delta$ interaction, Eq.~(\ref{dHijdg-1}) becomes
\begin{eqnarray} \label{dHijdg-2}
& & \mspace{-22mu} \lim_{g \rightarrow +\infty} \left( g^2 \langle \text{id}^{(g)} | \frac{d H}{d g} | P^{(g)} \rangle \right) = \frac{\hbar^4}{4 m^2} \sum_i \Bigl( \delta_{P, \text{id}} + \delta_{P, P_{i,i+1}} \Bigr) \nonumber \\
& & \mspace{-22mu} \times \int dz_1 \dotsi dz_N \, \delta (z_i-z_{i+1}) \bigl( D_i \psi_\text{id}^* \bigr) \bigl( D_i \psi_P \bigr)
\end{eqnarray}
with ${D_i \psi = D_i^+ \psi - D_i^- \psi}$ and
\begin{equation}
D_i^\pm \psi = \left( \frac{\partial}{\partial z_i} - \frac{\partial}{\partial z_{i+1}} \right) \psi \big|_{z_i=z_{i+1}\pm} .
\end{equation}
Note that we used ${\lim_{g \rightarrow +\infty} \psi_P^{(g)} = \psi_P}$ when we performed the limit. Using
\begin{equation}
\left. \frac{\partial \psi_F}{\partial z_{i+1}} \right|_{z_i=z_{i+1}} = - \left. \frac{\partial \psi_F}{\partial z_i} \right|_{z_i=z_{i+1}} ,
\end{equation}
\begin{equation}
A \bigr|_{z_i=z_{i+1}\pm} = (\pm 1) B
\end{equation}
with
\begin{equation}
B = \left. \frac{A}{\text{sgn}(z_i-z_{i+1})} \right|_{z_i=z_{i+1}} ,
\end{equation}
\begin{eqnarray}
& & \mspace{-60mu} \theta (z_1, \dotsc, z_N) \big|_{z_i=z_{i+1}+} \nonumber \\
& & \mspace{-60mu} = \theta (z_{P_{i,i+1}(1)}, \dotsc, z_{P_{i,i+1}(N)}) \big|_{z_i=z_{i+1}-} = 0 ,
\end{eqnarray}
and
\begin{eqnarray}
& & \mspace{-60mu} \theta (z_1, \dotsc, z_N) \big|_{z_i=z_{i+1}-} \nonumber \\
& & \mspace{-60mu} = \theta (z_{P_{i,i+1}(1)}, \dotsc, z_{P_{i,i+1}(N)}) \big|_{z_i=z_{i+1}+} \nonumber \\
& & \mspace{-60mu} = \theta (z_1, \dotsc, z_{i-1}, z_{i+1}, z_{i+1}, \dotsc, z_N) ,
\end{eqnarray}
we get
\begin{eqnarray}
& & \mspace{-22mu} D_i \psi_\text{id} = -D_i^- \psi_\text{id} = D_i \psi_{P_{i,i+1}} = D_i^+ \psi_{P_{i,i+1}} \nonumber \\
& & \mspace{-22mu} = 2 B \sqrt{N!} \theta (z_1, \dotsc, z_{i-1}, z_{i+1}, z_{i+1}, \dotsc, z_N) \left. \frac{\partial \psi_F}{\partial z_i} \right|_{z_i=z_{i+1}} \!\! . \nonumber \\
& &
\end{eqnarray}
Inserting this into Eq.~(\ref{dHijdg-2}) and using ${B^2 = 1}$ we get
\begin{equation} \label{dHijdg-3}
\mspace{-4mu} \frac{1}{g} \lim_{g \rightarrow +\infty} \left( g^2 \langle \text{id}^{(g)} | \frac{d H}{d g} | P^{(g)} \rangle \right) = \sum_i \Bigl( \delta_{P, \text{id}} + \delta_{P, P_{i,i+1}} \Bigr) J_i
\end{equation}
with
\begin{equation}
J_i = \frac{N! \hbar^4}{m^2 g} \int dz_1 \dotsi dz_N \delta (z_i-z_{i+1}) \theta (z_1, \dotsc, z_N) \left| \frac{\partial \psi_F}{\partial z_i} \right|^2 \! .
\end{equation}
Inserting Eq.~(\ref{dHijdg-3}) into Eq.~(\ref{Hij-3}) we get
\begin{equation} \label{Hij-4}
\langle \text{id}^{(g)} | H | P^{(g)} \rangle = \left( E_F - \sum_i J_i \right) \delta_{P, \text{id}} - \sum_i \delta_{P, P_{i,i+1}} J_i \, .
\end{equation}
Finally we insert this into Eq.~(\ref{Hij-2}) and obtain
\begin{eqnarray} \label{Hspin-2}
& & \mspace{-22mu} \langle m_1, \dotsc | \left( W_\pm^{(g)} \right)^\dagger H \, W_\pm^{(g)} | m_1', \dotsc \rangle \nonumber \\
& & \mspace{-22mu} = \langle m_1, \dotsc | \left[ \left( E_F - \sum_i J_i \right) \openone \pm \sum_i J_i P_{i, i+1} \right] | m_1', \dotsc \rangle \nonumber \\
& &
\end{eqnarray}
with ``$+$'' for fermions and ``$-$'' for bosons.

We would like to note that the effective interaction Hamiltonian~(\ref{Hspin-2}), which acts on many-body {\it spin} functions, originates from the tendency of the system to have a {\it spatial} wave function, which is most symmetric under the exchange of particles. The effective Hamiltonian following from Eq.~(\ref{Hij-4}), which acts on the {\it spatial} sector wave functions ${| P \rangle}$, has the form ${- \sum_i J_i P_{i, i+1}}$ (without diagonal terms). This Hamiltonian minimizes the energy of a pair of neighboring sector wave functions ${| P \rangle}$ and ${| P_{i, i+1} P \rangle}$, if it is in the symmetric superposition ${(| P \rangle + | P_{i, i+1} P \rangle) / \sqrt{2}}$, whereas the antisymmetric superposition ${(| P \rangle - | P_{i, i+1} P \rangle) / \sqrt{2}}$ maximizes the energy. Therefore, the system minimizes its energy, if as many as possible neighboring sector wave functions are in a symmetric superposition. This is in agreement with the theorem that the ground state of a system with the spin-independent Hamiltonian~(\ref{H}) strives to have as few as possible zero crossings of the spatial wave function~\cite{Lieb62, Eisenberg02}. We finally note that the effective interaction Hamiltonian~(\ref{Hspin-2}), like the original spin-independent Hamiltonian~(\ref{H}), commutes with the square of the total spin of the spin chain $\vec S^2$. This, together with the tendency of the system to have a most symmetric {\it spatial} wave function, leads in the case of spinful fermions to a ground state with minimal total spin~\cite{Lieb62}, whereas spinful bosons prefer a ground state with maximal total spin~\cite{Eisenberg02}.

\section{Exchange constants of the harmonic trap}
\label{app-J}

We compare in this appendix the nearest-neighbor exchange constants~(\ref{J}) of up to six harmonically trapped particles to its LDA approximations~(\ref{Japprox}). The TF profile of harmonically trapped noninteracting fermions is given by
\begin{equation}
n_\text{TF}(z) = \frac{1}{l \pi} \sqrt{2 N - \left( \frac{z}{l} \right)^2} .
\end{equation}
It is evaluated at the center-of-mass positions of the $i$th and ${(i+1)}$th particle,
\begin{equation}
z_i = \frac{1}{2} \int dz \, z \Bigl[ \rho^{(i)}(z) + \rho^{(i+1)}(z) \Bigr] .
\end{equation}
The particle densities ${\rho^{(i)}(z)}$ have been obtained from a fit to the exact total density. Table~\ref{table-J} shows the exact exchange constants of up to six harmonically trapped particles, obtained by computing the ${(N-1)}$-dimensional integrals of Eq.~(\ref{J}). The value in parentheses is the deviation of the LDA result~(\ref{Japprox}), which shows, as expected, an increasing agreement with increasing particle number.

\begin{table}
\begin{ruledtabular}
\begin{tabular}{cccc}
$N$ & $J_1 g / \left(\hbar^2 \omega^2 l \right)$ & $J_2 g / \left(\hbar^2 \omega^2 l \right)$ & $J_3 g / \left(\hbar^2 \omega^2 l \right)$ \\ \hline
$2$ & $\sqrt{\frac{\pi}{2}} = 0.797885$ ($6.4\%$)	&			&			\\
$3$ & $\frac{3^3}{2^3 \sqrt{2 \pi}} = 1.34643$ ($2.9\%$)&			&			\\
$4$ & $1.78765$ ($1.2\%$)				& $2.34651$ ($2.3\%$)	&			\\
$5$ & $2.16606$ ($0.14\%$)				& $3.17720$ ($1.6\%$)	&			\\
$6$ & $2.50218$ ($-0.55\%$)				& $3.90210$ ($1.1\%$)	& $4.35712$ ($1.2\%$)	\\
\end{tabular}
\end{ruledtabular}
\caption{\label{table-J} Nearest-neighbor exchange constants ${J_i g / \left(\hbar^2 \omega^2 l \right)}$ of ${N \leq 6}$ harmonically trapped particles. The value in parentheses is the deviation of the local density approximation. Note that ${J_{N-i} = J_i}$ due to the parity symmetry of the harmonic trap.}
\end{table}

\section{Validity regime of the spin-chain model}
\label{app-validity}

\subsection{Three spin-1/2 fermions}

\begin{figure}
\begin{center}
\includegraphics[width = \columnwidth]{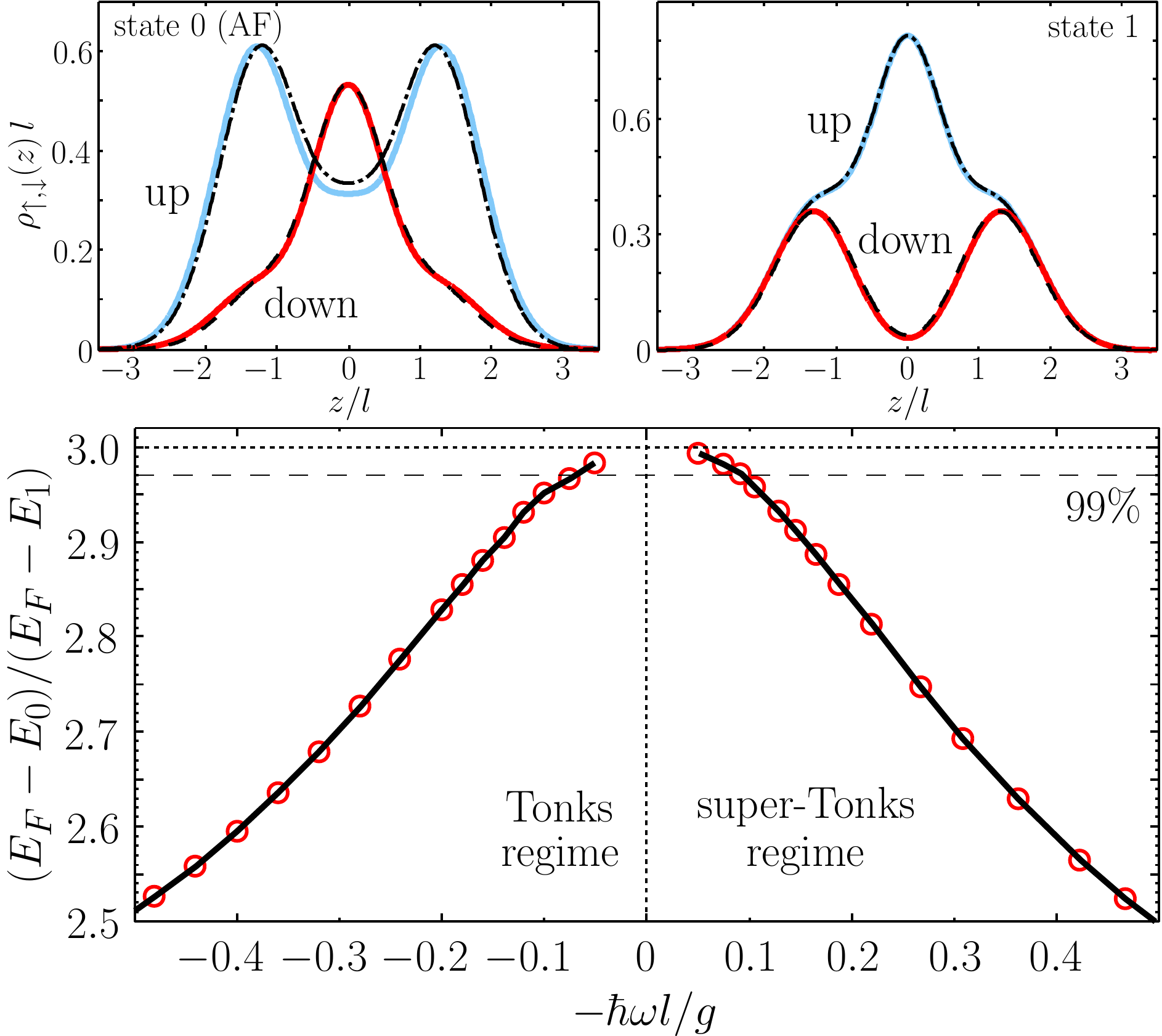}
\caption{(Color online) Comparison between the exact-diagonalization results of the original continuous model and the results obtained from the effective spin model for ${(N_\uparrow = 2, N_\downarrow = 1)}$ harmonically trapped spin-1/2 fermions. Top: Spin densities of the effective spin-chain model (solid lines) and of an exact diagonalization of the full Hamiltonian for ${g / (\hbar \omega l) = 10}$ in a harmonic trap (dashed and dash-dotted lines). Bottom: Ratio of energy differences of the ground-state multiplet as a function of ${-1/g}$ [solid (black) line with (red) circles]. The corresponding value of the spin-chain model is 3 (horizontal short-dashed line). The long-dashed line marks ${-1\%}$ deviation from this value.}
\label{fig-validity-uud}
\end{center}
\end{figure}

In this appendix, we compare analytically calculated energy differences and spin densities of ${(N_\uparrow = 2, N_\downarrow = 1)}$ harmonically trapped spin-1/2 fermions to those obtained by means of an exact diagonalization of the full Hamiltonian. We first calculate the spectrum and the eigenfunctions of the spin chain. Within the spin basis, ${| \uparrow, \uparrow, \downarrow \rangle}$, ${| \uparrow, \downarrow, \uparrow \rangle}$, and ${| \downarrow, \uparrow, \uparrow \rangle}$, the interaction Hamiltonian reads (note that ${J_1=J_2}$)
\begin{equation}
H_s = (E_F - 2 J_1) \openone + J_1
\begin{pmatrix}
  1 & 1 & 0 \\
  1 & 0 & 1 \\
  0 & 1 & 1
\end{pmatrix} .
\end{equation}
Its eigenstates are
\begin{equation}
| 0 \rangle = \frac{1}{\sqrt{6}} \Bigl( | \uparrow , \uparrow , \downarrow \rangle - 2 | \uparrow , \downarrow , \uparrow \rangle + | \downarrow , \uparrow , \uparrow \rangle \Bigr) ,
\end{equation}
\begin{equation}
| 1 \rangle = \frac{1}{\sqrt{2}} \Bigl( | \uparrow , \uparrow , \downarrow \rangle - | \downarrow , \uparrow , \uparrow \rangle \Bigr) ,
\end{equation}
and
\begin{equation}
| 2 \rangle = \frac{1}{\sqrt{3}} \Bigl( | \uparrow , \uparrow , \downarrow \rangle + | \uparrow , \downarrow , \uparrow \rangle + | \downarrow , \uparrow , \uparrow \rangle \Bigr) .
\end{equation}
They are simultaneously eigenstates of the square of the total spin $\vec S^2$ and the parity operator ${\Pi = -P_{1,3}}$,~\footnote{The parity operator $\Pi$, which acts originally on the sector wave functions in the usual way, ${(z_1, \dotsc, z_N) \rightarrow (-z_1, \dotsc, -z_N)}$, is transformed into the spin basis via the map~(\ref{map}), with the result ${\Pi = (\pm 1)^{\lfloor N/2 \rfloor} P_{1,N} P_{2,N-1} \dotsm}$ with ``$+$'' for bosons, ``$-$'' for fermions, and the common floor function ${\lfloor x \rfloor}$.} with eigenvalues ${S=1/2}$ and ${\Pi=-1}$ for ${| 0 \rangle}$, ${S=1/2}$ and ${\Pi=1}$ for ${| 1 \rangle}$, and ${S=3/2}$ and ${\Pi=-1}$ for ${| 2 \rangle}$~\cite{Harshman14}. The eigenenergies are ${E_0 = E_F - 3 J_1}$, ${E_1 = E_F - J_1}$, and ${E_2 = E_F}$. The ratio of energy differences is hence given by ${(E_F-E_0)/(E_F-E_1) = 3}$. Figure~\ref{fig-validity-uud}~(bottom) shows this ratio of energy differences as a function of ${-1/g}$. The result of an exact diagonalization of the full Hamiltonian of the harmonically trapped system [solid (black) line with (red) circles] approaches the analytical value 3, marked by the horizontal short-dashed line, in the (super-)Tonks regime. The deviation is smaller than 1\% for ${|\hbar \omega l / g| < 0.1}$.

Next we compare the density distributions obtained from the spin model with those resulting from the exact diagonalization of the original model. The analytical spin densities of the AF ground state ${| 0 \rangle}$ and the first excited state ${| 1 \rangle}$ are
\begin{equation}
\rho_\uparrow (z) = \frac{5}{6} \rho^{(1)} (z) + \frac{2}{6} \rho^{(2)} (z) + \frac{5}{6} \rho^{(3)} (z) ,
\end{equation}
\begin{equation}
\rho_\downarrow (z) = \frac{1}{6} \rho^{(1)} (z) + \frac{4}{6} \rho^{(2)} (z) + \frac{1}{6} \rho^{(3)} (z)
\end{equation}
and
\begin{equation}
\rho_\uparrow (z) = \frac{1}{2} \rho^{(1)} (z) + \rho^{(2)} (z) + \frac{1}{2} \rho^{(3)} (z) ,
\end{equation}
\begin{equation}
\rho_\downarrow (z) = \frac{1}{2} \rho^{(1)} (z) + \frac{1}{2} \rho^{(3)} (z) ,
\end{equation}
respectively [solid lines in Fig.~\ref{fig-validity-uud}~(top)]. The corresponding numerical results for ${g / (\hbar \omega l) = 10}$ (dashed and dash-dotted lines) agree very well with the analytical spin densities. The small deviation between the analytical and numerical spin densities of the AF ground state is larger than for the excited states (they agree for the F state ${| 2 \rangle}$).

\subsection{Four spin-1/2 fermions}

\begin{figure}
\begin{center}
\includegraphics[width = \columnwidth]{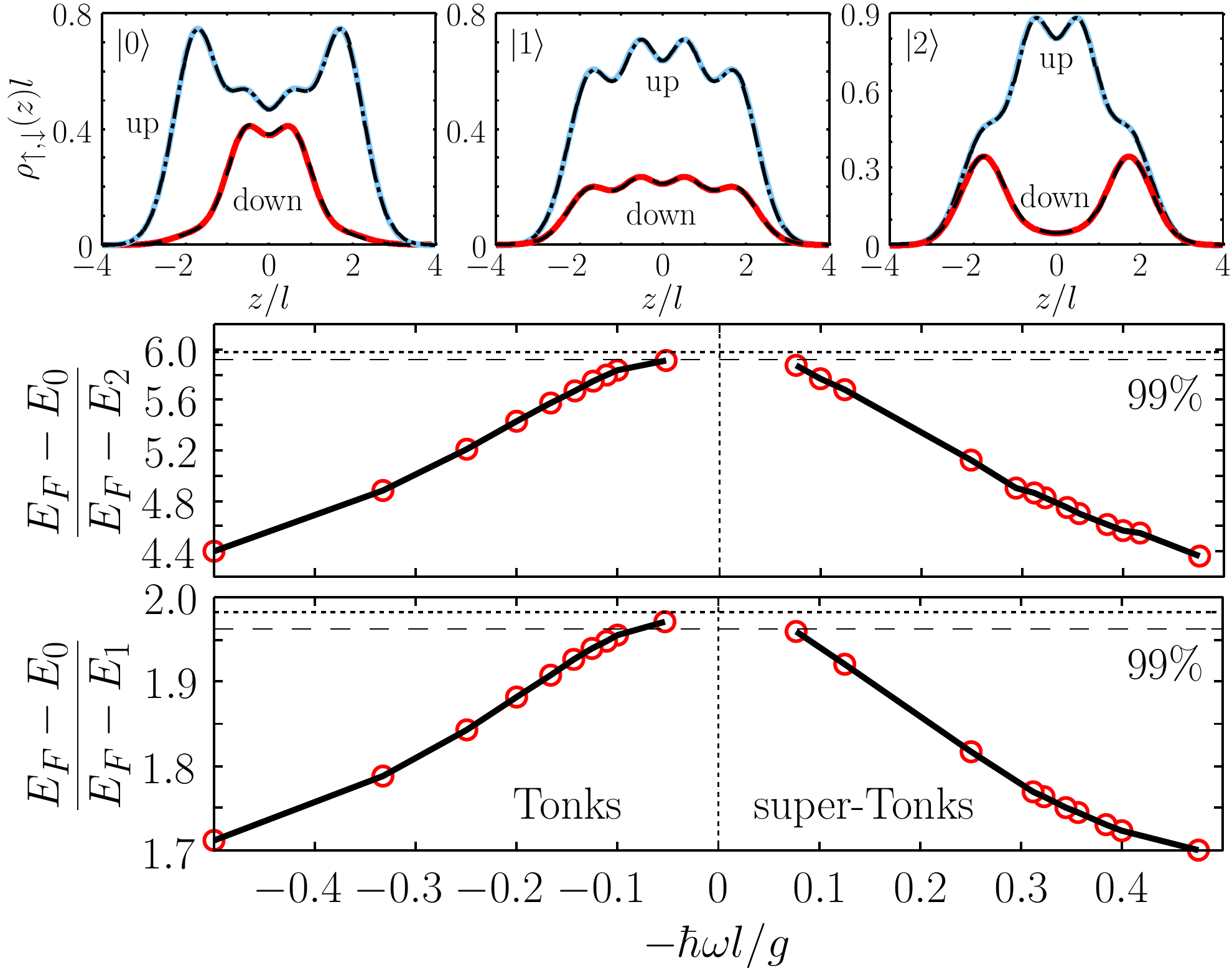}
\caption{(Color online) The same as Fig.~\ref{fig-validity-uud} for the ${(N_\uparrow = 3, N_\downarrow = 1)}$ system. Top: Spin densities of the effective spin-chain model (solid lines) and of an exact diagonalization of the full Hamiltonian for ${g / (\hbar \omega l) = 25}$ in a harmonic trap (dashed and dash-dotted lines). Bottom: Ratio of energy differences of the ground-state multiplet as a function of ${-1/g}$ [solid (black) line with (red) circles]. The corresponding values of the spin-chain model are $1.982$ and $5.983$, respectively (horizontal short-dashed lines). The horizontal long-dashed lines mark ${-1\%}$ deviation from these values.}
\label{fig-validity-uuud}
\end{center}
\end{figure}

Here we perform the same comparison as in the last section for four fermions in the ${(N_\uparrow = 3, N_\downarrow = 1)}$ configuration. Within the spin basis ${| \uparrow, \uparrow, \uparrow, \downarrow \rangle}$, ${| \uparrow, \uparrow, \downarrow, \uparrow \rangle}$, ${| \uparrow, \downarrow, \uparrow, \uparrow \rangle}$, and ${| \downarrow, \uparrow, \uparrow, \uparrow \rangle}$, the interaction Hamiltonian reads
\begin{equation} \label{Huuud}
H_s = (E_F - 2 J_1 - J_2) \openone +
\begin{pmatrix}
  J_1+J_2 & J_1 & 0 & 0 \\
  J_1 & J_1 & J_2 & 0 \\
  0 & J_2 & J_1 & J_1 \\
  0 & 0 & J_1 & J_1+J_2
\end{pmatrix} .
\end{equation}
With $J_1$ and $J_2$ of Table~\ref{table-J} the eigenenergies are given by
\begin{equation}
E_0 = E_F - J_1 - J_2 - \sqrt{J_1^2 + J_2^2} = E_F - 7.084 (\hbar \omega l / g) \hbar \omega ,
\end{equation}
\begin{equation}
E_1 = E_F - 2 J_1 = E_F - 3.575 (\hbar \omega l / g) \hbar \omega ,
\end{equation}
\begin{equation}
E_2 = E_F - J_1 - J_2 + \sqrt{J_1^2 + J_2^2} = E_F - 1.184 (\hbar \omega l / g) \hbar \omega ,
\end{equation}
and ${E_3 = E_F}$, which lead to the ratios of energy differences
\begin{equation}
\frac{E_F - E_0}{E_F - E_1} = \frac{J_1 + J_2 + \sqrt{J_1^2 + J_2^2}}{2 J_1} = 1.982
\end{equation}
and
\begin{equation}
\frac{E_F - E_0}{E_F - E_2} = \frac{J_1 + J_2 + \sqrt{J_1^2 + J_2^2}}{J_1 + J_2 - \sqrt{J_1^2 + J_2^2}} = 5.983 .
\end{equation}
We plot these ratios as a function of $-1/g$ in the bottom and middle panels of Fig.~\ref{fig-validity-uuud}. The results of an exact diagonalization of the full Hamiltonian of the harmonically trapped system [solid (black) lines with (red) circles] approach the results of the spin-chain model, $1.982$ and $5.983$, respectively (horizontal short-dashed lines), in the (super-)Tonks regime. Again, the deviation is only $\simeq 1\%$ for ${|\hbar \omega l / g| < 0.1}$.

The eigenstates of the spin Hamiltonian~(\ref{Huuud}) are
\begin{eqnarray}
|0\rangle & = & c_- \bigl( | \uparrow, \uparrow, \uparrow, \downarrow \rangle - | \downarrow, \uparrow, \uparrow, \uparrow \rangle \bigr) \nonumber \\
& & - c_+ \bigl( | \uparrow, \uparrow, \downarrow, \uparrow \rangle - | \uparrow, \downarrow, \uparrow, \uparrow \rangle \bigr) ,
\end{eqnarray}
\begin{equation}
|1\rangle = \frac{1}{2} \bigl( | \uparrow, \uparrow, \uparrow, \downarrow \rangle - | \uparrow, \uparrow, \downarrow, \uparrow \rangle - | \uparrow, \downarrow, \uparrow, \uparrow \rangle + | \downarrow, \uparrow, \uparrow, \uparrow \rangle \bigr) ,
\end{equation}
\begin{eqnarray}
|2\rangle & = & c_+ \bigl( | \uparrow, \uparrow, \uparrow, \downarrow \rangle - | \downarrow, \uparrow, \uparrow, \uparrow \rangle \bigr) \nonumber \\
& & + c_- \bigl( | \uparrow, \uparrow, \downarrow, \uparrow \rangle - | \uparrow, \downarrow, \uparrow, \uparrow \rangle \bigr) ,
\end{eqnarray}
and
\begin{equation}
|3\rangle = \frac{1}{2} \bigl( | \uparrow, \uparrow, \uparrow, \downarrow \rangle + | \uparrow, \uparrow, \downarrow, \uparrow \rangle + | \uparrow, \downarrow, \uparrow, \uparrow \rangle + | \downarrow, \uparrow, \uparrow, \uparrow \rangle \bigr)
\end{equation}
with
\begin{equation}
c_\pm = \frac{1}{2} \sqrt{1 \pm \frac{J_2}{\sqrt{J_1^2 + J_2^2}}} = \left\{
\begin{aligned}
& 0.6700 \; ( \text{for} \, +) , \\
& 0.2261 \; ( \text{for} \, -) .
\end{aligned}
\right.
\end{equation}
They are again eigenstates of $\vec S^2$ and ${\Pi = P_{1,4} P_{2,3}}$ with eigenvalues ${S=1}$ and ${\Pi=-1}$ for ${| 0 \rangle}$ and ${| 2 \rangle}$, ${S=1}$ and ${\Pi=1}$ for ${| 1 \rangle}$, and ${S=2}$ and ${\Pi=1}$ for ${| 3 \rangle}$~\cite{Harshman14}. The spin densities of the ground state ${|0\rangle}$ are
\begin{eqnarray}
\mspace{-40mu} \rho_\uparrow (z) & = & \bigl( |c_-|^2 + 2 |c_+|^2 \bigr) \left[ \rho^{(1)} (z) + \rho^{(4)} (z) \right] \nonumber \\
& & + \bigl( |c_+|^2 + 2 |c_-|^2 \bigr) \left[ \rho^{(2)} (z) + \rho^{(3)} (z) \right] ,
\end{eqnarray}
\begin{eqnarray}
\rho_\downarrow (z) & = & |c_-|^2 \left[ \rho^{(1)} (z) + \rho^{(4)} (z) \right] \nonumber \\
& & + |c_+|^2 \left[ \rho^{(2)} (z) + \rho^{(3)} (z) \right] ,
\end{eqnarray}
for the first excited state ${|1\rangle}$ we get
\begin{equation}
\rho_\uparrow (z) = \frac{3}{4} \left[ \rho^{(1)} (z) + \rho^{(2)} (z) + \rho^{(3)} (z) + \rho^{(4)} (z) \right] ,
\end{equation}
\begin{equation}
\rho_\downarrow (z) = \frac{1}{4} \left[ \rho^{(1)} (z) + \rho^{(2)} (z) + \rho^{(3)} (z) + \rho^{(4)} (z) \right] ,
\end{equation}
and for the second excited state ${|2\rangle}$ we get
\begin{eqnarray}
\mspace{-40mu} \rho_\uparrow (z) & = & \bigl( |c_+|^2 + 2 |c_-|^2 \bigr) \left[ \rho^{(1)} (z) + \rho^{(4)} (z) \right] \nonumber \\
& & + \bigl( |c_-|^2 + 2 |c_+|^2 \bigr) \left[ \rho^{(2)} (z) + \rho^{(3)} (z) \right] ,
\end{eqnarray}
\begin{eqnarray}
\rho_\downarrow (z) & = & |c_+|^2 \left[ \rho^{(1)} (z) + \rho^{(4)} (z) \right] \nonumber \\
& & + |c_-|^2 \left[ \rho^{(2)} (z) + \rho^{(3)} (z) \right] .
\end{eqnarray}
The spin densities of the spin-chain model, which are shown in the top row of Fig.~\ref{fig-validity-uuud} (solid lines), are compared to the numerical results for ${g / (\hbar \omega l) = 25}$ (dashed and dash-dotted lines) showing no visible difference.

\section{Gradient}
\label{app-gradient}

In this appendix, we transform the Hamiltonian of a $B$-field gradient into the spin basis. The matrix elements of a $B$-field gradient,
\begin{eqnarray} \label{VG}
& & V_G = (G/l) \sum_{i=1}^N \int dz_1 \dotsi dz_N | z_1, \dotsc, z_N \rangle \langle z_1, \dotsc, z_N | \nonumber \\
& & \mspace{140mu} \times z_i \sigma_z^{(i)} ,
\end{eqnarray}
are
\begin{eqnarray}
& & \langle m_1 , \dotsc | W_\pm^\dagger V_G W_\pm | m_1' , \dotsc \rangle \nonumber \\
& & = N! \langle m_1 , \dotsc | \langle \text{id} | S_\pm^\dagger V_G S_\pm | \text{id} \rangle | m_1' , \dotsc \rangle \nonumber \\
& & = \sum_P (\pm 1)^P \langle m_1 , \dotsc | \langle \text{id} | V_G | P \rangle | m_{P^{-1}(1)}' , \dotsc \rangle \nonumber \\
& & = \langle m_1 , \dotsc | \langle \text{id} | V_G | \text{id} \rangle | m_1' , \dotsc \rangle .
\end{eqnarray}
Here we used ${\langle \text{id} | V_G | P \rangle = \delta_{\text{id}, P} \langle \text{id} | V_G | \text{id} \rangle}$, which follows from the fact that different sectors of the many-body position space $\mathbb{R}^N$ have no overlap. Using Eq.~(\ref{VG}) we obtain
\begin{eqnarray}
& & \langle m_1 , \dotsc | W_\pm^\dagger V_G W_\pm | m_1' , \dotsc \rangle \nonumber \\
& & = \langle m_1 , \dotsc | \left[ (G/l) \sum_{i=1}^N \langle z \rangle_i \, \sigma_z^{(i)} \right] | m_1' , \dotsc \rangle ,
\end{eqnarray}
where ${\langle z \rangle_i}$ is the position of the $i$th spin,
\begin{eqnarray}
\langle z \rangle_i & = & \int dz_1 \dotsi dz_N z_i | \langle z_1, \dotsc, z_N | \text{id} \rangle |^2 \nonumber \\
& = & \int dz z \int dz_1 \dotsi dz_N \delta (z-z_i) | \langle z_1, \dotsc, z_N | \text{id} \rangle |^2 \nonumber \\
& = & \int dz z \rho^{(i)}(z) .
\end{eqnarray}

\section{Mean occupancies of harmonic-trap levels}
\label{app-occ}

\begin{figure}
\begin{center}
\includegraphics[width = \columnwidth]{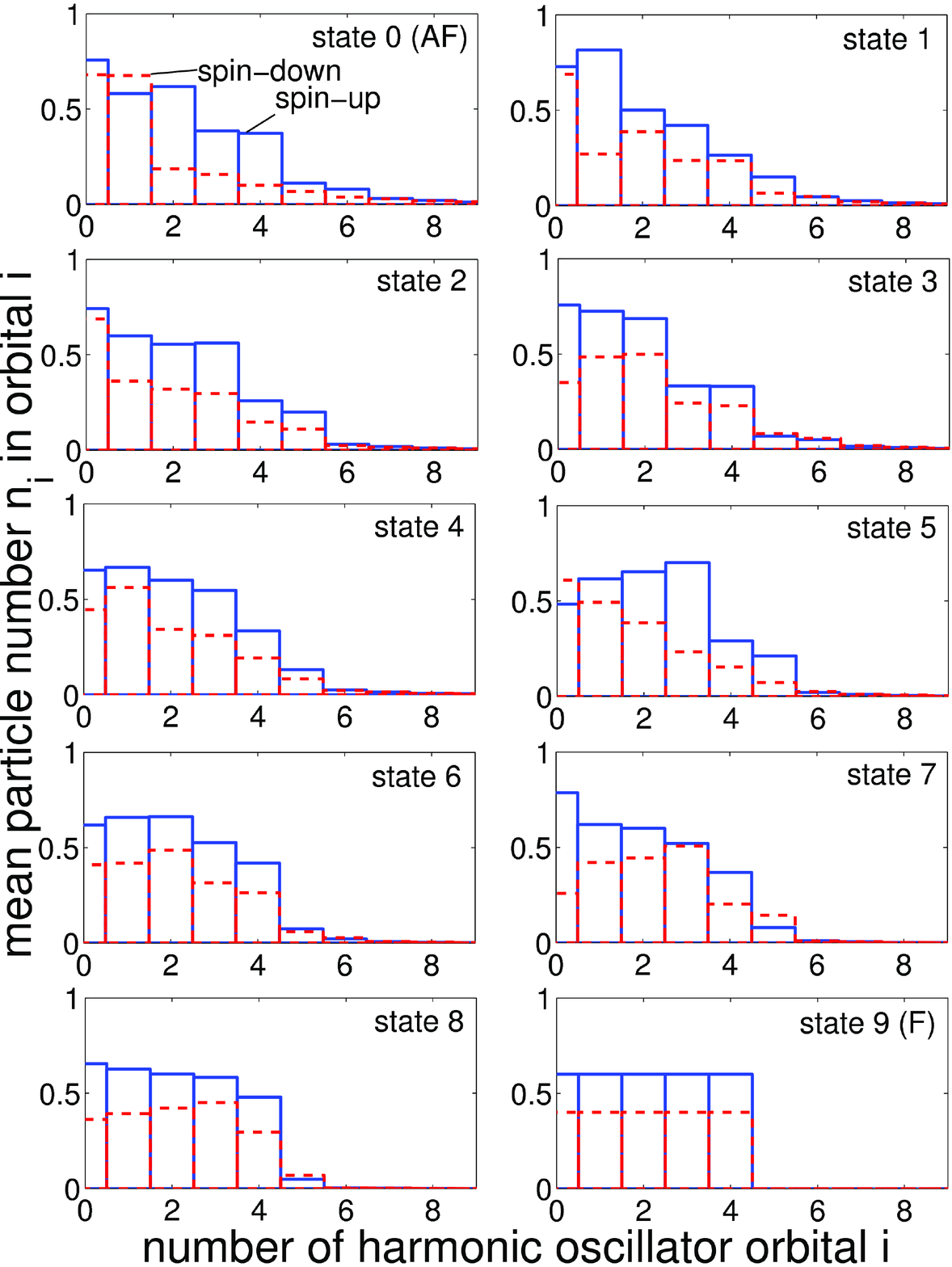}
\caption{(Color online) Mean occupation numbers ${\langle n_i \rangle}$ of the harmonic-oscillator orbitals of an ${(N_\uparrow = 3, N_\downarrow = 2)}$ Fermi system in the Tonks regime [${g / (\hbar \omega l) = 30}$, ground-state multiplet].}
\label{fig-occupancies-uuudd}
\end{center}
\end{figure}

In the main text, we mentioned that different states of the ground-state multiplet can be distinguished from each other by means of the mean occupancies ${\langle n_i \rangle}$ of the trap levels. Here we discuss this issue in more detail for a more involved example. Figure~\ref{fig-occupancies-uuudd} shows the mean occupancies ${\langle n_i \rangle}$ of the harmonic-oscillator orbitals of an ${(N_\uparrow = 3, N_\downarrow = 2)}$-fermion system in the Tonks regime. In this case, the ground-state multiplet consists of ten states. One sees that the AF state (state 0) and the F state (state 9) can be clearly distinguished from the others. The AF state features high occupancies of the lowest levels ${(n=0,1)}$ and small but nonzero occupancies in the ${n>4}$ levels. For the F state, the lowest five orbitals ${n=0,1,2,3,4}$ are equally populated while higher orbitals ${(n>4)}$ are empty.

In general, the occupation-number distribution is a measure of the symmetry of the spatial part of the many-body wave function. The AF state of spin-1/2 fermions has the most symmetric spatial wave function, which leads to high occupancies of the lowest levels ${(n=0,1)}$ and small but nonzero occupancies above the Fermi edge. The F state has a completely antisymmetric spatial wave function in which the states below the Fermi edge are equally populated while the states above the Fermi edge are empty. The other states interpolate between these extreme cases, i.e., the symmetry of the spatial wave functions decreases from state 0 to 9. The same arguments apply to the momentum distribution~\cite{Deuretzbacher08, Guan09, Fang11}.

\end{appendix}

\bibliographystyle{prsty}

\end{document}